\documentclass[aps,pra,reprint,amssymb,amsmath,superscriptaddress]{revtex4-2}
\usepackage{graphicx}
\usepackage{amsthm}
\usepackage{verbatim}

\def\be{\begin{equation}}
\def\ee{\end{equation}}
\def\ber{\begin{eqnarray}}
\def\eer{\end{eqnarray}}
\def\bern{\begin{eqnarray*}}
\def\eern{\end{eqnarray*}}

\def\dv{\mathbf{d}}
\def\rv{\mathbf{r}}

\def\jv{\mathbf{j}}

\def\qv{\mathbf{q}}

\def\bv{\mathbf{b}}
\def\Ev{\mathbf{E}}

\def\Fv{\mathbf{F}}
\def\Hv{\mathbf{H}}

\def\Rv{\mathbf{R}}

\def\Vv{\mathbf{V}}

\def\0v{\mathbf{0}}
\def\1v{\mathbf{1}}
\def\2v{\mathbf{2}}
\def\3v{\mathbf{3}}

\def\pa{\partial}

\DeclareMathAlphabet\mathbfcal{OMS}{cmsy}{b}{n}

\def\Im{ {\rm Im} \, }

\begin{document}

\title{Viscous current-induced forces}
\author{Vladimir~U.~Nazarov}
\affiliation{Fritz Haber Research Center of Molecular Dynamics, the Hebrew University of Jerusalem, Institute of Chemistry,  Israel}
\email{vladimir.nazarov@mail.huji.ac.il}
\author{Tchavdar~N. Todorov}
\affiliation{Queen's University Belfast, School of Mathematics and Physics, Belfast, UK}
\author{E.~K.~U.~Gross}
\affiliation{Fritz Haber Research Center of Molecular Dynamics, the Hebrew University of Jerusalem, Institute of Chemistry, Israel}

\begin{abstract}
We study the motion (translational, vibrational, and rotational) of a diatomic impurity immersed in an electron liquid and exposed to electronic current.
An approach based on the linear response time-dependent density functional theory combined with the Ehrenfest dynamics leads 
to a system of linear algebraic equations, which account for the competing and counteracting  effects of the current-induced force (electron wind) and the electronic friction. 
These forces, by means of the dynamic exchange-correlation kernel $f_{xc}(\rv,\rv',\omega)$, include the electronic viscosity contribution. 
Starting from the ground state at the equilibrium inter-nuclear distance  and applying a current pulse, we observe 
three phases of the motion: (I) acceleration due to the prevalence of the current-induced force, (II) stabilization upon balancing of the two forces,
and (III) deceleration due to the friction after the end of the pulse.
The viscous contribution to the force largely increases the acceleration (deceleration) at the first (third) phase of the process.
For the aluminium HEG electron density, we find this correction to
amount to up to 70\% of the total electron wind and friction effects. 
\end{abstract}

\maketitle

The current densities in atomic wires can exceed those in macroscopic conductors by many orders of magnitude \cite{Todorov-00}. Current flow in a lightbulb causes heating, light emission, and electromigration \cite{Sorbello-98}, leading to the eventual failure of the current-carrying element. 
So what should we expect in the microscopic world of atomic wires?

This question has prompted intense research into current-driven dynamics in nanoscale conductors for over 25 years, resulting in simulation techniques of great sophistication \cite{Lu-20,Stamenova-23}. Of central importance in these studies is the mean force exerted by the current on individual atomic nuclei. Techniques typically rely on static density functional theory (DFT) or self-consistent tight binding (TB) method for the calculation of this all-important quantity under non-equilibrium open-boundary conditions (see Ref.~\onlinecite{Dow-18} and references therein).

But it is known that static DFT/TB miss key dynamical effects, which have been shown to be of importance for related phenomena, such as electronic stopping \cite{Nazarov-07}, bulk impurities resistivity \cite{Nazarov-14-2}, and nanoscale conductance
\cite{ Koentopp-06,Sai-05,Jung-07,Sai-07}. These dynamical corrections, akin to electron viscosity, have never been investigated for current-induced forces, to the best of our knowledge. The aim of this Letter is to bridge this gap.

We develop a formalism based on the frequency-dependent kernel of time-dependent DFT (TDDFT) and apply it to the electron-wind force on impurities in jellium. The upshot is to correct the current-induced forces and the friction effect by up to 70\%. This viscous correction to the bare wind force is far from being of academic interest alone. When combined with thermal activation, this increase in the wind force can result in very significant changes to the impurity electromigration rates. A related phenomenon where these corrections become of central interest are non-conservative forces under current and the waterwheel effect \cite{Dundas-09}, an application we propose to study in the near future.  

We consider two classical nuclei, of atomic numbers $Z_1$ and $Z_2$, at positions $\Rv_1$ and $\Rv_2$, respectively, immersed in an otherwise homogeneous electron gas (HEG) of density parameter $r_s$, where $\bar {n}^{-1}=\frac{4}{3} \pi r_s^3$,  $\bar{n}$ being the HEG density
(atomic units are used throughout). 

{\em Equilibrium configuration}.--
In the ground state (GS) at the equilibrium inter-nuclear distance, the forces on each nucleus vanish 
\begin{equation}
\begin{split}
\Fv_\alpha & =
-\int \left[ \nabla_{\Rv_\alpha} \frac{Z_\alpha}{|\rv-\Rv_\alpha|} \right] n_0(\rv;\Rv_1,\Rv_2) d\rv  \\
& -\nabla_{\Rv_\alpha} \frac{Z_\alpha Z_\beta}{|\Rv_\alpha-\Rv_\beta|}= \0v, \, \alpha=1,2, \, \beta=2,1,
\label{equi}
\end{split}
\end{equation}
where $n_0(\rv;\Rv_1,\Rv_2)$ is the nuclear-position-dependent GS electronic density.

{\em Time-dependent perturbation}.--
We apply to the equilibrium state a weak external electric field 
\begin{equation}
\delta \Ev_{ext}(t)=\delta \Ev_{ext}(\omega) e^{-i \omega t}.
\label{Eext}
\end{equation}
To first order in the perturbation (\ref{Eext}), the current density induced in the system is
\begin{equation}
\begin{split}
\delta j_{ind,i}(\rv,\omega) &=  \frac{c}{i\omega} \int  \hat{\chi}_{i j}(\rv,\rv',\omega) 
\left[ \delta E_{ext,j}(\omega) \right.\\
&\left. +\sum\limits_{\gamma=1}^{2} (\delta \Rv_\gamma(\omega)\cdot \nabla') \nabla'_j \frac{Z_\gamma  }{|\rv'-\Rv_\gamma|} \right]d\rv',
\end{split}
\label{j}
\end{equation}
where $i$ and $j$ are Cartesian indices, and summation over the doubly repeated index $j$ is implied.
In Eq.~(\ref{j}),  $\hat{\chi}_{i j}(\rv,\rv',\omega)$ is the tensorial {\em current} density response function \cite{Vignale-98}, in which a parametric dependence on $\Rv_{1,2}$ is implied. $\delta \Rv_\gamma(\omega)$ is the displacement, to first order in $\Ev_{ext}(\omega$), of nucleus $\gamma$ from its equilibrium position. The two terms in the square brackets in Eq.~(\ref{j}) stand for the bare external field and that of the Coulomb charges of the displaced nuclei, respectively.

Based on Eq.~(\ref{j}), in Appendix \ref{SecI} we show that the electric field due to the
dynamical redistribution of the electrons is 
\begin{equation}
\begin{split}
&\delta \Ev_e(\rv,\omega) 
\! =  \! \frac{\omega^2}{\omega^2 \! - \! \omega_p^2} \delta \Ev_{ext}(\omega)
\! + \! \frac{4\pi c}{ \omega^2} \nabla^{-2} \nabla \nabla_i \! \int  \! \hat{\chi}_{i j}(\rv,\rv',\omega) \\
&\times \left[ \delta E_{ext,j}(\omega) +\sum\limits_{\gamma=1}^{2} (\delta \Rv_\gamma(\omega)\cdot \nabla') \nabla'_j \frac{Z_\gamma  }{|\rv'-\Rv_\gamma|} \right]d\rv' ,
\end{split}
\label{intm}
\end{equation}
where $\omega_p=\sqrt{4 \pi \bar{n}}$ is the plasma frequency of the HEG. 

As the next step,
using a relation between the tensorial $\hat{\chi}$ of the TDCDFT and the scalar $\chi$ of the TDDFT,
we rewrite Eq.~(\ref{intm}) as (see Appendix \ref{SecII})
\begin{widetext}
\begin{equation}
\begin{split}
\delta \Ev_e(\rv,\omega) 
&=  \frac{\omega^2}{\omega^2-\omega_p^2} \delta \Ev_{ext}(\omega) +    \frac{1}{\omega^2-\omega_p^2} 
 \nabla    \int \frac{1}{|\rv-\rv''|}
[ \chi({\bf r}'',{\bf r}',\omega)- \chi({\bf r}'',{\bf r}',0)] (\delta \Ev_{ext}(\omega)\cdot \nabla')    V_0({\bf r}')
\, d {\bf r}' d\rv''  \\
&+  \nabla \int   \frac{1}{|\rv-\rv''|}  \chi({\bf r}'',{\bf r}',\omega)
\left[  \sum\limits_{\gamma=1}^{2} (\delta \Rv_\gamma(\omega)\cdot \nabla') \frac{Z_\gamma  }{|\rv'-\Rv_\gamma|} \right] d\rv' d\rv'',
\end{split}
\label{intm4}
\end{equation}
\end{widetext}
where
\begin{equation}
V_0(\rv)=-\sum\limits_{\gamma=1}^2 \frac{Z_\gamma}{|\rv-\Rv_\gamma|}
\label{V0}
\end{equation}
is the bare static potential by the nuclei. The advantage of Eq.~(\ref{intm4}) over Eq.~(\ref{intm}) is that the former relies on the more
tractable scalar density response function $\chi$.

The last step is to evaluate the forces on each nucleus. To the first order in the perturbation,  they read
\begin{equation}
\begin{split}
&\delta \Fv_\alpha(\omega) = Z_\alpha \delta \Ev_e(\Rv_\alpha,\omega) \\ &-Z_\alpha [(\delta \Rv_\alpha(\omega)\cdot \nabla_{\Rv_\alpha}) +(\delta \Rv_\beta(\omega)\cdot \nabla_{\Rv_\beta})  ] \nabla_{\Rv_\alpha} \frac{Z_\beta}{|\Rv_\alpha -\Rv_\beta|}   \\
&+Z_\alpha (\delta \Rv_\alpha(\omega)\cdot \nabla_{\Rv_\alpha}) \nabla_{\Rv_\alpha} \int \frac{n_0(\rv)}{|\Rv_\alpha-\rv|} d\rv, \ \beta\ne\alpha.
\end{split}
\label{force}
\end{equation}
Indeed, the first term on the RHS of Eq.~(\ref{force}) is the field of Eq.~(\ref{intm4}) at the {\em unshifted} position of the $\alpha$-th nucleus times its charge.
The second line stands for the change of the bare force from another nucleus due to the displacements of both of them.
The third line accounts for the change of the force in the potential of the ground-state electronic distribution due to the displacement of the nucleus $\alpha$. 
Equations (\ref{intm4}) - (\ref{force}) constitute the main result of our theory. On the equal footing, they include current-induced forces and the electronic friction \cite{Dow-17,Dow-18}, and the electron-electron interaction is fully accounted for through, the formally exact, density response function $\chi(\rv,\rv',\omega)$ \cite{Gross-85}.

Within the Ehrenfest dynamics, Eqs.~(\ref{intm4}) - (\ref{force}) are  complemented with the Newton's equations
\begin{equation}
-M_\alpha \omega^2 \delta \Rv_\alpha(\omega)=\delta  \Fv_\alpha(\omega),
\end{equation}
thus closing the system of equations to be solved to find  the displacements $\delta \Rv_\alpha(\omega)$.

The most challenging element of the described scheme is the determination of the density response function $\chi(\rv,\rv',\omega; \Rv_1,\Rv_2)$ of the system of the HEG plus two nuclei in their respective  positions  (here we indicate the parametric dependence on the latter explicitly).
In order to avoid significant computational difficulties while still keeping the essential physics, in the implementation of our theory we resort to the 

{\em Weak electrons-impurity interaction approximation}.--
This approximation amounts to replacing the density response function in Eq.~(\ref{intm4}) by its HEG counterpart $\chi^h(|\rv-\rv'|,\omega)$ while, in the equilibrium condition (\ref{equi}), taking
\begin{equation}
n_0(\rv; \Rv_1,\Rv_2)= \int \chi^h(|\rv-\rv'|,0) V_0(\rv') d\rv'.
\label{n0weak}
\end{equation}
Then the force and the potential energy at the given separation $d$ between the nuclei are evaluated to (see Appendix \ref{SecIII})
\begin{equation}
F  \! = \! \frac{Z_1 Z_2 }{d^2} \! \left[
8  \! \int  \! d q \frac{\chi^h(q,0)}{q^3} 
 (\sin q d \! - \!  q d \cos q d)  \! + \! 1 \right],
\label{F_weak_equil}
\end{equation}
\begin{equation}
U  = \frac{Z_1 Z_2}{d} \left[
8  \int  d q\frac{\chi^h(q,0)}{q^3} 
 \sin  q d   + 1 \right],
\label{U_weak_equil}
\end{equation}
where we have Fourier transformed $\chi^h(\rv,0)$ to the wave-vector variable $\qv$.
\begin{figure}[h!]
\includegraphics[width=\columnwidth, trim=68 0 15 0, clip=true] {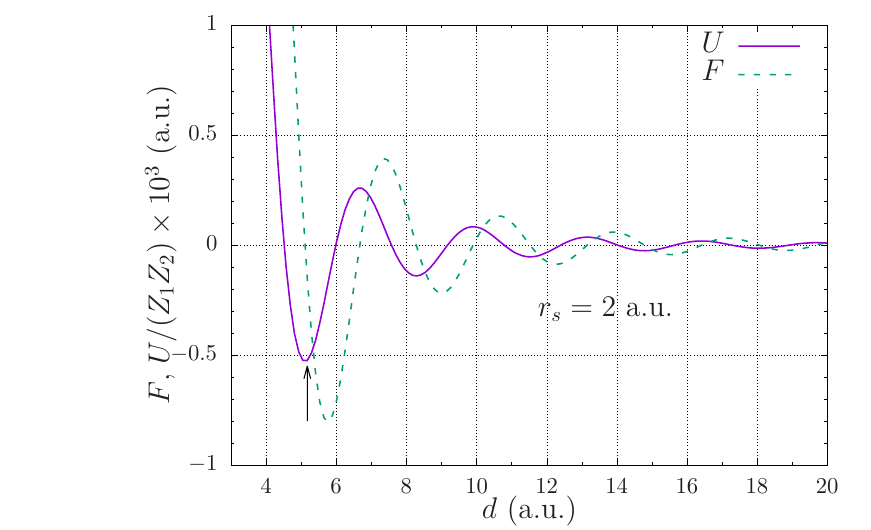}
\caption{\label{stat} The potential energy $U$ (\ref{U_weak_equil}) and the force $F$ (\ref{F_weak_equil}) versus the distance $d$ between two point charges $Z_1$ and $Z_2$ in the HEG of the density parameter $r_s=2$ a.u. The equilibrium distance between the charges (marked with an arrow) is $d=5.17$ a.u.
$\chi^h(q,0)$ used was obtained with Eq.~\eqref{GK} and the MCP07 approximation for $f_{xc}(q,0)$.\cite{Ruzsinszky-20} }
\end{figure}
In Fig.~\ref{stat}, the force and the potential energy of Eqs.~(\ref{F_weak_equil}) and (\ref{U_weak_equil})
are plotted versus the inter-nuclear  separation $d$,  for HEG of density parameter $r_s=2$.
We note that, in the weak interaction approximation, the equilibrium separation is independent on the nuclear charges, being a function, via $\chi^h$, of the HEG density only.

Instead of the external field $\delta \Ev_{ext}(\omega)$, we introduce the current density $\overline{\delta \jv} (\omega)$ in the HEG as it would be in the absence of the impurity, the two quantities being related by 
\begin{equation}
\overline{\delta \jv} (\omega)=  \frac{i \omega \omega_p^2}{4\pi (\omega^2-\omega_p^2)} \delta \Ev_{ext}(\omega), 
\end{equation}
which is a consequence of the Drude formula. 
We, furthermore, introduce the coordinate of the center of mass (c.m.)
\begin{equation*}
\delta \Rv_c(\omega)= \frac{M_1 \delta \Rv_1(\omega)+ M_2 \delta \Rv_1(\omega)}{M_c}, \ M_c=M_1+M_2,
\end{equation*}
the relative coordinate
\begin{equation}
\delta \Rv_r(\omega)=\delta \Rv_2(\omega)-\delta \Rv_1(\omega),
\end{equation}
and the corresponding velocities $\delta \Vv_c(\omega)$ and $\delta \Vv_r(\omega)$, 
which, finally, leads to a system of coupled equations of motion in terms of the latter velocities
\begin{widetext}
\begin{equation}
\begin{split}
-\omega   \delta \Vv_c(\omega)
& \! = \!  - \frac{4\pi  (Z_1 \! + \! Z_2) \omega}{M_c \omega_p^2}  \overline{\delta \jv} (\omega)  
+   \frac{2}{\pi M_c\omega} \int d\qv  \frac{\qv}{q^4}  \left[ \chi^h(q,\omega)-\chi^h(q,0)\right] \left[Z_1^2+Z_2^2  +2 Z_1 Z_2  e^{-i \qv\cdot \dv } \right]  
\left\{\left[\frac{\overline{\delta \jv}(\omega)}{\bar{n}}-\delta \Vv _c(\omega)\right]\cdot \qv\right\}   \\
&-   \frac{2}{\pi M_c^2 \omega} \! \int  \! d\qv \frac{\qv}{q^4}  \left[ \chi^h(q,\omega)-\chi^h(q,0)\right] 
\left[  M_1 Z_2^2-M_2 Z_1^2   +Z_1 Z_2 (M_1-M_2) e^{-i \qv\cdot \dv } \right] ( \delta \Vv_r(\omega)\cdot \qv),  
\end{split}
\label{vcr1}
\end{equation}
\begin{equation}
\begin{split}
-\omega  \delta \Vv_r(\omega)
&=  - \frac{4\pi   \omega}{  \omega_p^2}  \left( \frac{Z_2}{M_2}-\frac{Z_1}{M_1} \right) \overline{\delta \jv} (\omega)  \\
&-   \frac{2}{\pi M_c \omega} \int d\qv \frac{\qv}{q^4}  \left[ \chi^h(q,\omega)-\chi^h(q,0)\right] (  \delta \Vv _r(\omega)\cdot \qv) 
\left[ \frac{M_1 Z_2^2}{M_2 } + \frac{M_2 Z_1^2}{M_1 }   - 2 Z_1 Z_2 e^{i \qv\cdot \dv } \right]  \\
&+   \frac{2}{\pi \omega} \int  d\qv \frac{\qv}{q^4}  \left[ \chi^h(q,\omega)-\chi^h(q,0)\right] \left\{\left[\frac{\overline{\delta \jv}(\omega)}{\bar{n}}-\delta \Vv _c(\omega)\right]\cdot \qv\right\}
\left[\frac{Z_2^2}{M_2} -\frac{Z_1^2}{M_1} + \left(\frac{1}{M_2}-\frac{1}{M_1} \right)  Z_1 Z_2 e^{i \qv\cdot \dv } \right] \\
&+    \frac{2}{\pi \omega} \left(\frac{1}{M_1}+\frac{1}{M_2} \right) \int  d\qv  \frac{\qv}{q^4} Z_1 Z_2 \chi^h(q,0) e^{i \qv\cdot \dv }
 [\delta \Vv_r(\omega)\cdot \qv]     
- \frac{1}{ \omega} \left(\frac{1}{M_1}+ \frac{1}{M_2}\right) [ \delta \Vv_r(\omega)\cdot \nabla_{\dv}] \nabla_{\dv} \frac{Z_1  Z_2}{d}  .
\end{split}
\label{vcr2}
\end{equation}
\end{widetext}

While the c.m. motion  and the relative one are coupled in Eqs.~(\ref{vcr1})-(\ref{vcr2}),
we show that the motion in the direction parallel to $\dv$ and that in the perpendicular direction are independent (see Appendix \ref{SecIV})
This allows us to study the two geometries separately.
Individual terms in Eqs.~(\ref{vcr1})-(\ref{vcr2}) can be attributed a meaning as follows.
The first terms on the RHS of both equations are due to the direct field, i.e., the external field (\ref{Eext})
screened in HEG. The second terms stand for the combined action of the electron wind and the friction, in Eq.~(\ref{vcr1}), and the friction only in Eq.~(\ref{vcr2}).
In the limit $\omega\to 0$, a factor $\pa \chi(q,\omega)/\pa \omega|_{\omega=0}$ appears under the integrals. The latter is familiar from theories
of the stopping power for ions and the impurity resistivity \cite{Nazarov-07,Nazarov-14-2}.
At finite $\omega$, these terms still keep their meaning, although they are quantitatively changed according to the magnitude of $\omega$.
The third terms are the cross ones, accounting for the c.m. motion dependence on the relative one and vice verse. 
The last two terms in Eq.~(\ref{vcr2}) stand for the elastic restoring force in the relative motion, as becomes clear noting that
$i\delta \Vv_r(\omega)/\omega =\delta \Rv_r(\omega)$.
 This force gives rise to vibrations in the system shown in Fig 4. When the motion is decomposed into the parallel and perpendicular to $\dv$ components, this force survives in the parallel part only [see Eqs.~\eqref{ppar} and \eqref{pperp} of Appendix \ref{SecIV}].

Equations (\ref{vcr1})-(\ref{vcr2}) solve our problem in the case of a monochromatic field (\ref{Eext}). Furthermore,  utilizing the linearity of the present approach, by means of the forward and inverse Fourier transforms,
we can construct the solution for an arbitrary current pulse $\bar{\delta \jv}(t)$. 
We choose a pulse of rectangular shape
\begin{equation}
\bar{\delta \jv}(t)=[H(T-t)-H(-t)] \, \jv_0 ,
\label{jrect}
\end{equation}
where $H(t)$ is the Heaviside step function and $T$ is the duration of the pulse.

\begin{figure}[h!]
\includegraphics[width=\columnwidth, trim=28 0 15 0, clip=true] {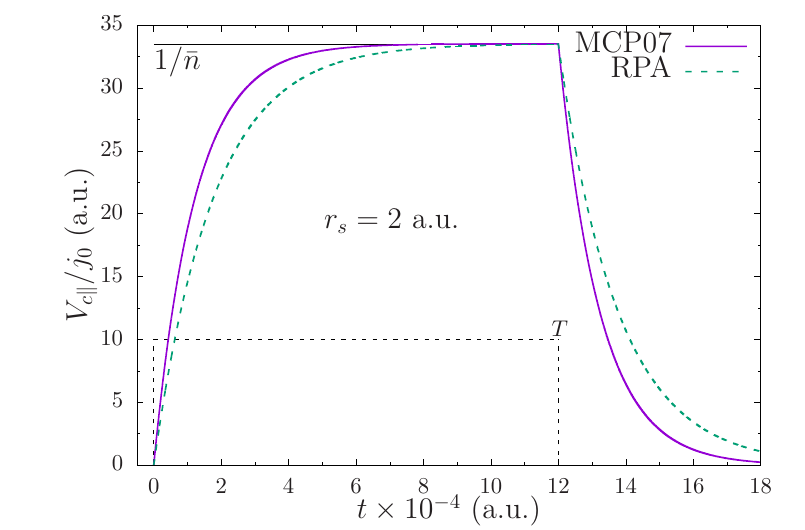}
\caption{\label{rectparc} 
Velocity of the center-of-mass  of the impurity comprised of a proton and deuteron in HEG of $r_s=2$ a.u. A rectangular current pulse is applied in the direction {\it parallel} to the impurity's  axis. At $t<0$, the system is in its ground state, at the equilibrium separation between the nuclei.
The solid (MCP07) and dashed (RPA) lines show results of calculations by Eqs.~(\ref{vcr1}) -- (\ref{vcr2}) with the use of $f_{xc}^h(q,\omega)$ of Ref.~\onlinecite{Ruzsinszky-20} and the random phase approximation [$f_{xc}^h(q,\omega)=0$], respectively.
The pulse is schematically shown with a rectangle in short-dashed lines.}
\end{figure}

\begin{figure}[h!]
\includegraphics[width=\columnwidth, trim=28 0 15 0, clip=true]{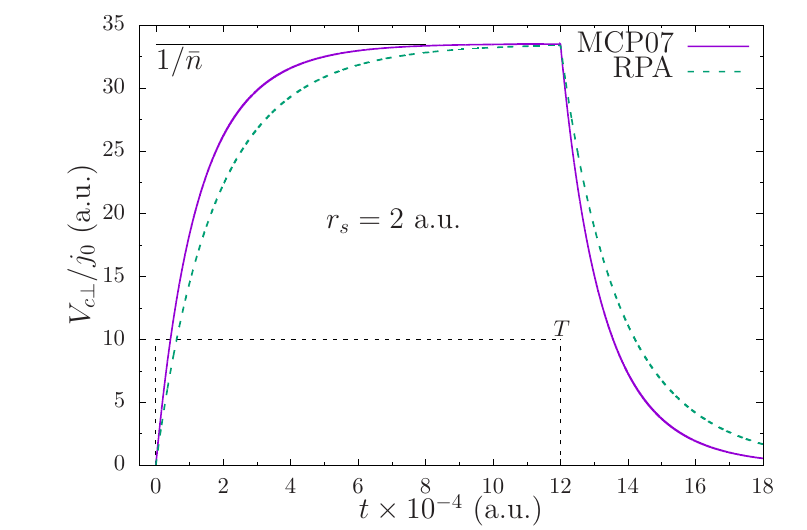}
\caption{\label{rectperpc} 
Same as Fig.~\ref{rectparc}, but the pulse is applied in the direction {\it perpendicular} to the impurity axis.}
\end{figure}

\begin{figure}[h!]
\includegraphics[width=\columnwidth, trim=28 0 15 0, clip=true]{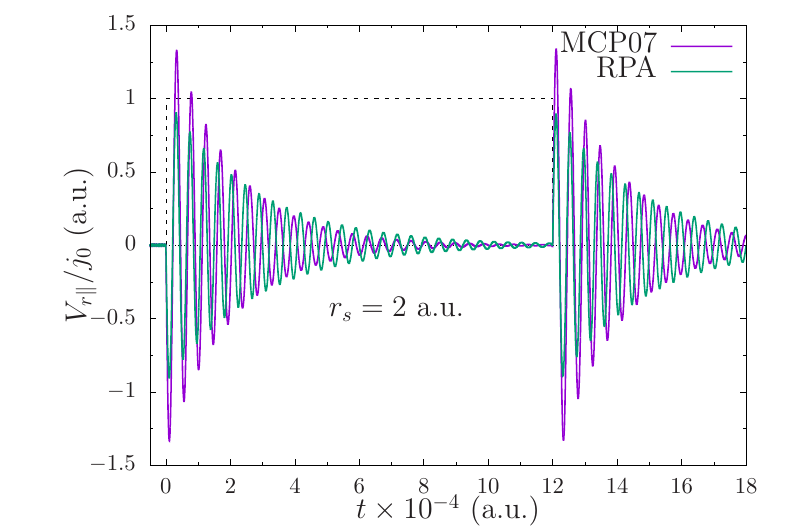}
\caption{\label{rectparr} 
Same as Fig.~\ref{rectparc}, but for the velocity of the nuclei relative to each other. }
\end{figure}

\begin{figure}[h!]
\includegraphics[width=\columnwidth, trim=28 0 15 0, clip=true]{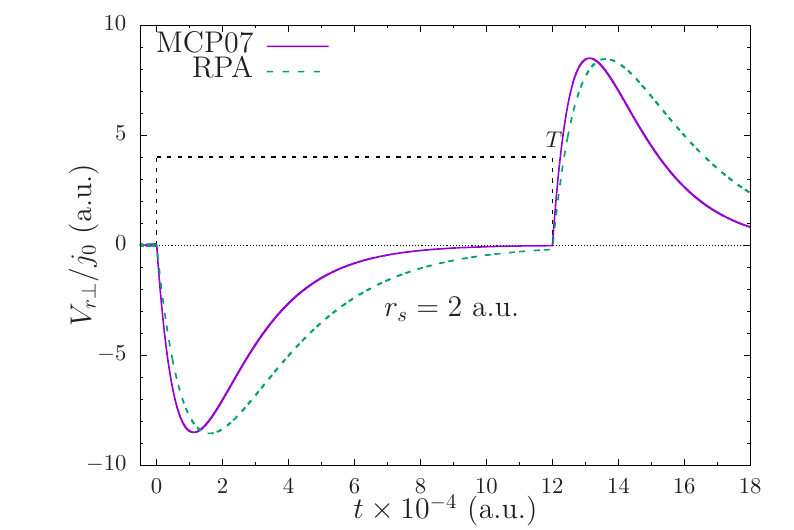}
\caption{\label{rectperpr} 
Same as Fig.~\ref{rectparr}, but the pulse is applied in the direction {\it perpendicular} to the impurity's  axis.}
\end{figure}

We have conducted calculations using Eqs.~(\ref{vcr1})-(\ref{jrect}) for the impurity comprised of a proton and a deuteron in a HEG of $r_s=2$. 
In Figs.~\ref{rectparc} and \ref{rectperpc}, we present the time evolution of the velocity of the c.m. of the impurity,
in the direction parallel to the axis $\dv$ and perpendicular to it, respectively.
In Figs.~\ref{rectparr} - \ref{rectperpr} the same is shown for the relative velocities.
The c.m. motion (Figs.~\ref{rectparc} and \ref{rectperpc}) are qualitatively similar in the two geometries.
Upon the application of the pulse at $t=0$, acceleration of the impurity as a whole under the electron wind takes place.
The acceleration gradually slows down as the counteracting friction force grows with the increase of the velocity, until the c.m. velocity stabilizes at a value of $\jv_0/\bar{n}$, i.e., at the electronic drift velocity.
Upon the pulse switching off at $t=T$, the electron wind force disappears, leading to a fast slowing down of the impurity under the friction force alone.

In Fig.~\ref{rectparr}, the relative velocity evolution in the direction parallel to the impurity axis is shown.
A dominating feature here is the vibrational motion around the equilibrium separation between the two nuclei. The vibrations are, however, of attenuating amplitudes, after the kicks of the pulse switching on and off. This attenuation is due to the coupling between the relative and c.m. motions in Eqs.~~(\ref{vcr1})-(\ref{vcr2}):
when the velocity of the c.m. approaches saturation around the middle of the pulse (see Fig.~\ref{rectparc}), the electron wind ceases to support the vibrations, while the friction persists, leading to the decay of the oscillations. The frequency and the damping of the oscillations are worked out in
Appendix \ref{SecV}.

Figure~\ref{rectperpr} also shows the evolution of the relative velocity, but in the geometry with the current  perpendicular to the impurity axis.
This regime corresponds to the rotation of the impurity around its c.m.
Similar to vibrations, the velocity of the rotation quickly increases after the pulse switching on and off, but it falls off in the region of the stabilization (see Appendix \ref{SecV}).

In our theory, a factor of the primary role is the coupling  between the motion of the c.m. of the impurity and that of the nuclei relative to each other. This is due to the mediation by the medium and, obviously, it is absent in the motion in vacuum.
One of the consequences of the coupling is that,
in the middle of the pulse duration, when the stabilization of the c.m. velocity is reached, the DC current does not support vibrations and rotation, because the impurity is moving as a whole with the saturation velocity $\jv_0/\bar{n}$. However,  the pulse switching on or off constitutes kicks on the system, as a result all frequencies get involved, leading to the  commencement/resumption of vibrations and rotation, which die out afterwards due to the friction
(see Figs.~\ref{rectparr} and \ref{rectperpr}).


In the determination of  $\chi^h(q,\omega)$ in Eqs.~(\ref{vcr1}) -- (\ref{vcr2}),
we rely on the linear response TDDFT equality \cite{Gross-85}
\begin{equation}
\frac{1}{\chi^h(q,\omega)}=\frac{1}{\chi_s^h(q,\omega)}-\frac{4\pi}{q^2}-f_{xc}^h(q,\omega),
\label{GK}
\end{equation} 
where $\chi_s^h(q,\omega)$ is the Kohn-Sham single particle density response function and $f_{xc}^h(q,\omega)$ is the dynamic xc kernel.
While the former is known exactly and analytically by the Lindhard formula \cite{Giuliani&Vignale}, 
the knowledge of the latter is limited to approximations.
In our calculations we use the constraint-based $f_{xc}^h(q,\omega)$ of Ref.~\onlinecite{Ruzsinszky-20} (MCP07). 
Results are compared with those in the random phase  approximation (RPA) [i.e., $f_{xc}^h(q,\omega)=0$].

From Figs.~\ref{rectparc} -- \ref{rectperpr} we conclude that the electronic viscosity, i.e., the frequency-dependence of $f_{xc}$, plays an important role in the motion.
In the c.m. motion (Figs.~ \ref{rectparc} and \ref{rectperpc}), it leads to an increase in the acceleration due the electron wind at the first phase of the motion, and to a faster deceleration upon the end of the pulse, the latter being due to friction.
The maximal effect of the viscosity can be observed in the case of the rotational motion (Fig.~\ref{rectperpr}), where, in the stabilization region (phase II), it leads to a decrease of the velocity by up to 70\%.

In conclusion, we have studied the motion of a diatomic impurity in an electron liquid under the action of an electronic current.
A consistent linear response TDDFT approach combined with Ehrenfest dynamics has been utilized.
In contrast to applications of this method to the related problems of the electronic stopping power and the impurity resistivity in metals,
we find the role of the viscosity of the electron liquid to be of major importance. In particular, in the case of the rotational motion, the viscosity contribution comprises up to 70\% of the electron wind and the electronic friction effects.
Fundamental implications for the electromigration theory are envisaged. 

\acknowledgements
This project has received funding from the European Research Council (ERC) under the European Union's Horizon 2020 research and information programme (Grant Agreement No. ERC-2017-AdG-788890).


%

\appendix

\onecolumngrid

\section{Derivation of Eq.~(\ref{intm}) }

\label{SecI}

\noindent
By Maxwell's equations
\begin{align}
&\nabla \times \delta \Ev(\rv,\omega)=\frac{i\omega}{c} \delta \Hv(\rv,\omega), \\
&\nabla\times \delta \Hv(\rv,\omega)= - \frac{i \omega}{c} \delta \Ev(\rv,\omega) -\frac{4\pi}{c} \delta \jv(\rv,\omega), \label{22}
\end{align}
we can write
\begin{equation}
\delta \Ev(\rv,\omega)=-\frac{4\pi}{i\omega} \delta \jv(\rv,\omega) +\frac{c^2}{\omega^2} \nabla\times[\nabla\times \delta \Ev(\rv,\omega)].
\label{1}
\end{equation}
Since, for a uniform field,
\begin{equation}
\Ev_{ext}(\omega)=-\frac{4\pi}{i\omega} \jv_{ext}(\omega),
\end{equation}
we can rewrite Eq.~(\ref{1}) as
\begin{equation}
\delta \Ev(\rv,\omega)=\delta \Ev_{ext}(\omega)-\frac{4\pi}{i\omega} \delta \jv_{ind}(\rv,\omega) +\frac{c^2}{\omega^2} \nabla\times[\nabla\times \delta \Ev(\rv,\omega)].
\label{2}
\end{equation}
where
\begin{equation}
\jv_{ind}(\rv,\omega)=\jv(\rv,\omega)-\jv_{ext}(\omega).
\end{equation}
Using a vector analysis identity, Eq.~\ref{2} can be written as
\begin{equation}
\delta \Ev(\rv,\omega)= \delta \Ev_{ext}(\omega)-\frac{4\pi}{i\omega} \delta \jv_{ind}(\rv,\omega) +\frac{c^2}{\omega^2} \nabla [\nabla\cdot \delta \Ev(\rv,\omega)]
-\frac{c^2}{\omega^2} \nabla^2 \delta \Ev(\rv,\omega).
\end{equation}
or, with account of Eq.~(\ref{22}),
\begin{equation}
\delta \Ev(\rv,\omega)=\delta \Ev_{ext}(\omega)-\frac{4\pi}{i\omega} \delta \jv_{ind}(\rv,\omega) -\frac{4\pi c^2}{i \omega^3} \nabla [\nabla\cdot \delta \jv_{ind}(\rv,\omega)]
-\frac{c^2}{\omega^2} \nabla^2 \delta \Ev(\rv,\omega).
\end{equation}
Adding and subtracting $-\frac{4\pi}{i\omega} \overline{\delta \jv}_{ind}(\omega)$, we have
\begin{equation}
\delta \Ev(\rv,\omega)+\frac{c^2}{\omega^2} \nabla^2 \delta \Ev(\rv,\omega) = \delta \Ev_{ext}(\omega)-\frac{4\pi}{i\omega} \overline{\delta \jv}_{ind}(\omega)-\frac{4\pi}{i\omega} [\delta \jv_{ind}(\rv,\omega)-\overline{\delta \jv}_{ind}(\omega)] -\frac{4\pi c^2}{i \omega^3} \nabla [\nabla\cdot \delta \jv_{ind}(\rv,\omega)].
\end{equation}
and then
\begin{equation}
\delta \Ev(\rv,\omega) = \delta \Ev_{ext}(\omega)-\frac{4\pi}{i\omega} \overline{\delta \jv}_{ind}(\omega)-(1+\frac{c^2}{\omega^2} \nabla^2)^{-1} \left[\frac{4\pi}{i\omega} [\delta \jv_{ind}(\rv,\omega)-\overline{\delta \jv}_{ind}(\omega)] +\frac{4\pi c^2}{i \omega^3} \nabla [\nabla\cdot \delta \jv_{ind}(\rv,\omega)]\right],
\end{equation}
where we have used the fact that, on a uniform field, $(1+\frac{c^2}{\omega^2} \nabla^2)^{-1}$ acts as the identity operator.
Then, in the limit $c\to \infty$,
\begin{equation}
\delta \Ev(\rv,\omega) = \delta \Ev_{ext}(\omega)-\frac{4\pi}{i\omega} \overline{\delta \jv}_{ind}(\omega)-  \frac{4\pi}{i \omega} \nabla^{-2} \nabla [\nabla\cdot \delta \jv_{ind}(\rv,\omega)] .
\end{equation}
Since quantities averaged over the whole space are those of the HEG, we have
\begin{equation}
\begin{split}
\delta \Ev(\rv,\omega) &= \delta \Ev_{ext}(\omega)+\overline{\delta \Ev}_{ind}(\omega)-  \frac{4\pi}{i \omega} \nabla^{-2} \nabla [\nabla\cdot \delta \jv_{ind}(\rv,\omega)]\\
&= \overline{\delta \Ev}(\omega)-  \frac{4\pi}{i \omega} \nabla^{-2} \nabla [\nabla\cdot \delta \jv_{ind}(\rv,\omega)]
=  \frac{1}{1-\frac{\omega_p^2}{\omega^2}} \delta \Ev_{ext}(\omega)-  \frac{4\pi}{i \omega} \nabla^{-2} \nabla [\nabla\cdot \delta \jv_{ind}(\rv,\omega)] .
\end{split}
\label{4}
\end{equation}
Equation (\ref{4}) together with Eq.~(\ref{j}) lead us to Eq.~(\ref{intm}).

\section{Derivation of Eq.~(\ref{intm4})}
\label{SecII}

Using the relation \cite{Nazarov-14-2}
\begin{equation}
\begin{split}
   \left( \omega^2- \omega_p^2\right) \int
\hat{\chi}_{ij}({\bf r},{\bf r}',\omega) \, d {\bf r}'=   
 \int
\hat{\chi}_{ik}({\bf r},{\bf r}',\omega) \, \nabla'_k \nabla'_j V_0({\bf r}')
\, d {\bf r}' +  \frac{\omega^2}{c} n_0({\bf r})\,\delta_{ij},
\end{split}
\label{SRS}
\end{equation}
$V_0(\rv)$ given by Eq.~(\ref{V0}),
we have from Eq.~(\ref{intm})
\begin{equation}
\begin{split}
\delta \Ev(\rv,\omega) 
&=  \frac{1}{1-\frac{\omega_p^2}{\omega^2}} \delta \Ev_{ext}(\omega)\\
&+  \frac{4\pi c}{ \omega^2 (\omega^2-\omega_p^2)} \nabla^{-2} \nabla\left[ \int
 \nabla_i  \hat{\chi}_{ik}({\bf r},{\bf r}',\omega) \nabla'_k \nabla'_j  \,  V_0({\bf r}')
\, d {\bf r}' +  \frac{\omega^2}{c}  \nabla_i  n_0({\bf r})\,\delta_{ij} 
  \right] \delta E_{ext,j}(\omega) \\
&+  \frac{4\pi c}{ \omega^2} \nabla^{-2} \nabla\int   \nabla_i  \hat{\chi}_{i j}(\rv,\rv',\omega) \nabla'_j 
\left[  \sum\limits_{\gamma=1}^{2} (\delta \Rv_\gamma(\omega)\cdot \nabla') \frac{Z_\gamma  }{|\rv'-\Rv_\gamma|} \right]d\rv' .
\label{intm2}
\end{split}
\end{equation}
With the use of the relation between the tensorial $\hat{\chi}$ of the TDCDFT and the scalar one $\chi$ of the TDDFT \cite{Nazarov-14-2}
\begin{equation}
\chi({\bf r},{\bf r}',\omega) =  -\frac{c}{ \omega^2} \nabla_i \hat{\chi}_{ij}({\bf r},{\bf r}',\omega)  \nabla'_j,
\end{equation}
Eq.~(\ref{intm2}) can be rewritten as
\begin{equation}
\begin{split}
\delta \Ev(\rv,\omega) 
&=  \frac{1}{1-\frac{\omega_p^2}{\omega^2}} \delta \Ev_{ext}(\omega)
-  \frac{4\pi}{\omega^2-\omega_p^2}  \nabla  \nabla^{-2} \left[ \int
 \chi({\bf r},{\bf r}',\omega) (\delta \Ev_{ext}(\omega)\cdot \nabla')    V_0({\bf r}')
\, d {\bf r}' -  (\delta \Ev_{ext}(\omega) \cdot \nabla)  n_0({\bf r})
  \right] \\
&- 4\pi  \nabla \nabla^{-2} \int   \chi({\bf r},{\bf r}',\omega)
\left[  \sum\limits_{\gamma=1}^{2} (\delta \Rv_\gamma(\omega)\cdot \nabla') \frac{Z_\gamma  }{|\rv'-\Rv_\gamma|} \right]d\rv' ,
\end{split}
\label{intm3}
\end{equation}
or, with the use of the static sum-rule,
\begin{equation}
\int \chi(\rv,\rv',0) \nabla' V_0(\rv') d\rv' = \nabla n_0(\rv),
\end{equation}
Eq.~(\ref{intm3}) can be written as
\begin{equation}
\begin{split}
\delta \Ev(\rv,\omega) 
&=  \frac{1}{1-\frac{\omega_p^2}{\omega^2}} \delta \Ev_{ext}(\omega)
-  \frac{4\pi}{\omega^2-\omega_p^2}  \nabla  \nabla^{-2}  \int
[ \chi({\bf r},{\bf r}',\omega)- \chi({\bf r},{\bf r}',0)] ( \delta \Ev_{ext}(\omega)\cdot \nabla')    V_0({\bf r}')
\, d {\bf r}'  \\
&- 4\pi  \nabla \nabla^{-2} \int   \chi({\bf r},{\bf r}',\omega)
\left[  \sum\limits_{\gamma=1}^{2} (\delta \Rv_\gamma(\omega)\cdot \nabla') \frac{Z_\gamma  }{|\rv'-\Rv_\gamma|} \right]d\rv',
\end{split}
\end{equation}
which proves Eq.~(\ref{intm4}).

\section{Derivation of Eqs.~(\ref{F_weak_equil}) and (\ref{U_weak_equil})}
\label{SecIII}

Rewriting the force of Eq.~(\ref{equi}) with the substitution of $n_0(\rv)$ in the weak interaction regime (\ref{n0weak}), we obtain
\begin{equation}
\begin{split}
\Fv_\alpha & = 
\frac{1}{(2\pi)^3} \int \left[ \nabla_\rv \frac{Z_\alpha}{|\rv-\Rv_\alpha|} \right] 
\chi^h(q,0) e^{i (\rv-\rv')\cdot \qv} \left[ \frac{Z_\alpha}{|\rv'-\Rv_\alpha|} + \frac{Z_\beta}{|\rv'-\Rv_\beta|} \right] d\rv d\rv' d\qv -\nabla_{\Rv_\alpha} \frac{Z_\alpha Z_\beta}{|\Rv_\alpha-\Rv_\beta|}, \, \alpha=1,2, \, \beta=2,1,
\end{split}
\end{equation}
or, after the integration by parts,
\begin{equation}
\begin{split}
\Fv_\alpha & = 
-\frac{1}{(2\pi)^3} \int i \qv\frac{Z_\alpha}{|\rv-\Rv_\alpha|}
\chi^h(q,0) e^{i (\rv-\rv')\cdot \qv} \left[ \frac{Z_\alpha}{|\rv'-\Rv_\alpha|} + \frac{Z_\beta}{|\rv'-\Rv_\beta|} \right] d\rv d\rv' d\qv -\nabla_{\Rv_\alpha} \frac{Z_\alpha Z_\beta}{|\Rv_\alpha-\Rv_\beta|}, \, \alpha=1,2, \, \beta=2,1.
\end{split}
\end{equation}
or
\begin{equation}
\begin{split}
\Fv_\alpha & = 
-\frac{1}{(2\pi)^3} \int i \qv\frac{Z_\alpha}{|\rv|}
\chi^h(q,0) e^{i (\rv-\rv')\cdot \qv} \frac{Z_\alpha}{|\rv'|} d\rv d\rv' d\qv \\
&-\frac{1}{(2\pi)^3} \int i \qv\frac{Z_\alpha}{|\rv-\Rv_\alpha|}
\chi^h(q,0) e^{i (\rv-\rv')\cdot \qv} \frac{Z_\beta}{|\rv'-\Rv_\beta|}  d\rv d\rv' d\qv -\nabla_{\Rv_\alpha} \frac{Z_\alpha Z_\beta}{|\Rv_\alpha-\Rv_\beta|}, \, \alpha=1,2, \, \beta=2,1,
\end{split}
\end{equation}
where the integration variables substitutions has been performed in the first term. Obviously, the latter evaluates to zero, and we have
\begin{equation}
\begin{split}
\Fv_\alpha & = 
-\frac{1}{(2\pi)^3} \int i \qv\frac{Z_\alpha}{|\rv-\Rv_\alpha|}
\chi^h(q,0) e^{i (\rv-\rv')\cdot \qv} \frac{Z_\beta}{|\rv'-\Rv_\beta|}  d\rv d\rv' d\qv -\nabla_{\Rv_\alpha} \frac{Z_\alpha Z_\beta}{|\Rv_\alpha-\Rv_\beta|}, \, \alpha=1,2, \, \beta=2,1,
\end{split}
\end{equation}
In the last equation, integrations over $\rv$ and $\rv'$ are carried out explicitly, producing
\begin{equation}
\begin{split}
\Fv_\alpha & = -
\frac{(4\pi)^2 Z_\alpha Z_\beta}{(2\pi)^3} \int i \frac{\qv}{q^4} 
\chi^h(q,0) e^{-i \dv\cdot \qv}  d\qv -\nabla_{\dv} \frac{Z_\alpha Z_\beta}{d}= \0v, \, \alpha=1,2, \, \beta=2,1.
\end{split}
\label{11}
\end{equation}

Equation (\ref{F_weak_equil}) is retrieved by the integration in Eq.~(\ref{11}) over angular variables of the vector $\qv$.
Finally, Eq.~(\ref{U_weak_equil}) is obtained by the formula
\begin{equation}
U(d)=\int\limits_d^\infty F(d) d d.
\end{equation}

\section{Proof of the independence of the motions parallel and perpendicular to the impurity's axis within the weak interaction approximation} 
\label{SecIV}

For an arbitrary vector $\bv=\bv_\|+\bv_\perp$, where $\bv_\|$ and $\bv_\perp$ are vectors parallel and perpendicular to $\dv$, respectively, equalities hold
\begin{align}
 &\int \qv (\bv\cdot \qv)  d \Omega_\qv = \frac{4 \pi}{3} q^2 \bv, \label{b1}\\
& \int \qv (\bv_\|\cdot \qv) e^{i \dv \cdot \qv} d \Omega_\qv =
 \frac{4 \pi}{d^3 q}  \bv_\|  [2 q d \cos q d+(q^2 d^2-2) \sin q d ], \\
& \int \qv (\bv_\perp\cdot \qv) e^{i \dv \cdot \qv} d \Omega_\qv =
  \frac{4\pi}{d^3 q}\bv_\perp (\sin q d - q d \cos q d ),
  \label{b3}
\end{align}
where the integrations are taken over the full solid angle of the vector $\qv$. In Eqs.~(\ref{vcr1}) and (\ref{vcr2}), let us decompose
\begin{align}
&\overline{\delta \jv} (\omega) =\overline{\delta \jv}_\| (\omega) +\overline{\delta \jv}_\perp (\omega) , \\
&\delta \Vv_c(\omega)=\delta \Vv_{c\|}(\omega)+\delta \Vv_{c\perp}(\omega), \\
&\delta \Vv_r(\omega)= \delta \Vv_{r\|}(\omega)+\delta \Vv_{c\perp}(\omega).
\end{align}
Then,
with the use of Eqs.~(\ref{b1})-(\ref{b3}), 
\begin{equation}
\begin{split}
&-\omega M_c  \delta \Vv_{c\|}(\omega)
 \! = \!  - \frac{4\pi  (Z_1 \! + \! Z_2) \omega}{ \omega_p^2}  \overline{\delta \jv}_\| (\omega)  \\
& + \!  \frac{8}{ \omega} \int  d q    \left[ \chi^h(q,\omega) \! - \! \chi^h(q,0)\right] 
\left\{\frac{ 1}{3} (Z_1^2+Z_2^2)  + \frac{2 Z_1 Z_2}{q^3 d^3}   [2 q d \cos q d+(q^2 d^2-2) \sin q d ] \right\}  
\left[\frac{\overline{\delta \jv}_\|(\omega)}{\bar{n}} \! - \! \delta \Vv _{c\|}(\omega)\right]    \\
&-   \frac{8}{ M_c \omega} \! \int  \!  d q   \left[ \chi^h(q,\omega)-\chi^h(q,0)\right] 
\left\{ \frac{1}{3}  (M_1 Z_2^2-M_2 Z_1^2)   + (M_1-M_2) \frac{Z_1 Z_2}{q^3 d^3}    [2 q d \cos q d+(q^2 d^2-2) \sin q d ]\right\}  \delta \Vv_{r\|}(\omega)   .  
\end{split}
\label{vcr1par}
\end{equation}
\begin{equation}
\begin{split}
&-\omega  \delta \Vv_{r\|}(\omega)
=  - \frac{4\pi   \omega}{  \omega_p^2}  \left( \frac{Z_2}{M_2}-\frac{Z_1}{M_1} \right) \overline{\delta \jv}_\| (\omega)  \\
&+   \frac{8}{ \omega} \int  d q   \left[ \chi^h(q,\omega)-\chi^h(q,0)\right] 
\left\{\frac{1}{3} \left(\frac{Z_2^2}{M_2} -\frac{Z_1^2}{M_1}\right) + \left(\frac{1}{M_2}-\frac{1}{M_1} \right)  \frac{Z_1 Z_2 }{q^3 d^3}  [2 q d \cos q d+(q^2 d^2-2) \sin q d ] \right\} \left[\frac{\overline{\delta \jv}_\|(\omega)}{\bar{n}}-\delta \Vv _{c\|}(\omega)\right]\\
&-   \frac{8}{ M_c \omega} \int d q   \left[ \chi^h(q,\omega)-\chi^h(q,0)\right]   
\left[  \frac{1}{3}  \left(\frac{M_1 Z_2^2}{M_2 } + \frac{M_2 Z_1^2}{M_1 } \right)  - 2 \frac{Z_1 Z_2 }{q^3 d^3}    [2 q d \cos q d+(q^2 d^2-2) \sin q d ]\right] \delta \Vv _{r\|}(\omega)  \\
&+    \frac{8}{ \omega} \left(\frac{1}{M_1}+\frac{1}{M_2} \right) \int  d q  \chi^h(q,0) \frac{Z_1 Z_2}{q^3 d^3} [2 q d \cos q d+(q^2 d^2-2) \sin q d ]
 \delta \Vv_{r\|}(\omega) 
- \frac{2 Z_1  Z_2}{ \omega d^3} \left(\frac{1}{M_1}+ \frac{1}{M_2}\right) \delta \Vv_{r_\|}(\omega).
\end{split}
\label{vcr2par}
\end{equation}

\begin{equation}
\begin{split}
-\omega M_c  \delta \Vv_{c\perp}(\omega)
& \! = \!  - \frac{4\pi  (Z_1 \! + \! Z_2) \omega}{ \omega_p^2}  \overline{\delta \jv}_\perp (\omega)  
\! + \!   \frac{8}{ \omega} \! \int \! \! d q  \!  \left[ \chi^h(q,\omega) \! - \! \chi^h(q,0)\right] \!  \! \left[\frac{1}{3} (Z_1^2+Z_2^2) \!  + \! \frac{2 Z_1 Z_2}{q^3 d^3} (\sin q d \! -  \! q d \cos q d ) \right]  \! \!
\left[\frac{\overline{\delta \jv}_\perp(\omega)}{\bar{n}} \! - \! \delta \Vv _{c\perp}(\omega)\right]   \\
&-   \frac{8}{ M_c \omega} \! \int  \!  d q   \left[ \chi^h(q,\omega)-\chi^h(q,0)\right] 
\left[ \frac{1}{3}  ( M_1 Z_2^2-M_2 Z_1^2 )  + (M_1-M_2) \frac{Z_1 Z_2}{q^3 d^3} (\sin q d - q d \cos q d ) \right]  \delta \Vv_{r\perp}(\omega),  
\end{split}
\label{vcr1perp}
\end{equation}
\begin{equation}
\begin{split}
-\omega  \delta \Vv_{r\perp}(\omega)
&=  - \frac{4\pi   \omega}{  \omega_p^2}  \left( \frac{Z_2}{M_2}-\frac{Z_1}{M_1} \right) \overline{\delta \jv}_\perp (\omega)  \\
&+   \frac{8}{ \omega} \int  d q   \left[ \chi^h(q,\omega)-\chi^h(q,0)\right] 
\left[\frac{1}{3}  \left(\frac{Z_2^2}{M_2} -\frac{Z_1^2}{M_1}\right) + \left(\frac{1}{M_2}-\frac{1}{M_1} \right)  \frac{Z_1 Z_2 }{q^3 d^3} (\sin q d - q d \cos q d )\right] \left[\frac{\overline{\delta \jv}_\perp(\omega)}{\bar{n}}-\delta \Vv _{c\perp}(\omega)\right]\\
&-   \frac{8}{ M_c \omega} \int d q   \left[ \chi^h(q,\omega)-\chi^h(q,0)\right]  
\left[\frac{1}{3} \left( \frac{M_1 Z_2^2}{M_2 } + \frac{M_2 Z_1^2}{M_1 }\right)   - 2 \frac{Z_1 Z_2 }{q^3 d^3} (\sin q d - q d \cos q d ) \right] \delta \Vv _{r\perp}(\omega) \\
&+    \frac{8}{ \omega} \left(\frac{1}{M_1}+\frac{1}{M_2} \right) \int  d q  Z_1 Z_2 \chi^h(q,0) \frac{1}{q^3 d^3} (\sin q d - q d \cos q d )
 \delta \Vv_{r\perp}(\omega)
+ \frac{Z_1  Z_2}{ \omega d^3} \left(\frac{1}{M_1}+ \frac{1}{M_2}\right)  \delta \Vv_{r_\perp}(\omega).
\end{split}
\label{vcr2perp}
\end{equation}

Finally, we note that, because $d$ is the equilibrium inter-nuclear distance nullifying the force of  Eq.~(\ref{F_weak_equil}),
Eqs.~(\ref{vcr2par}) and (\ref{vcr2perp}) can be rewritten as
\begin{equation}
\begin{split}
&-\omega  \delta \Vv_{r\|}(\omega)
=  - \frac{4\pi   \omega}{  \omega_p^2}  \left( \frac{Z_2}{M_2}-\frac{Z_1}{M_1} \right) \overline{\delta \jv}_\| (\omega)  \\
&+   \frac{8}{ \omega} \int  d q   \left[ \chi^h(q,\omega)-\chi^h(q,0)\right] 
\left\{\frac{1}{3} \left(\frac{Z_2^2}{M_2} -\frac{Z_1^2}{M_1}\right) + \left(\frac{1}{M_2}-\frac{1}{M_1} \right)  \frac{Z_1 Z_2 }{q^3 d^3}  [2 q d \cos q d+(q^2 d^2-2) \sin q d ] \right\} \left[\frac{\overline{\delta \jv}_\|(\omega)}{\bar{n}}-\delta \Vv _{c\|}(\omega)\right]\\
&-   \frac{8}{ M_c \omega} \int d q   \left[ \chi^h(q,\omega)-\chi^h(q,0)\right]   
\left[  \frac{1}{3}  \left(\frac{M_1 Z_2^2}{M_2 } + \frac{M_2 Z_1^2}{M_1 } \right)  - 2 \frac{Z_1 Z_2 }{q^3 d^3}    [2 q d \cos q d+(q^2 d^2-2) \sin q d ]\right] \delta \Vv _{r\|}(\omega)  \\
&+    \frac{8}{ \omega} \left(\frac{1}{M_1}+\frac{1}{M_2} \right) \int  d q  \chi^h(q,0) \sin q d \frac{Z_1 Z_2}{q d} 
 \delta \Vv_{r\|}(\omega) .
\end{split}
\tag{\ref{vcr2par}$'$}
\label{ppar}
\end{equation}
\begin{equation}
\begin{split}
-\omega  \delta \Vv_{r\perp}(\omega)
&=  - \frac{4\pi   \omega}{  \omega_p^2}  \left( \frac{Z_2}{M_2}-\frac{Z_1}{M_1} \right) \overline{\delta \jv}_\perp (\omega)  \\
&+   \frac{8}{ \omega} \int  d q   \left[ \chi^h(q,\omega)-\chi^h(q,0)\right] 
\left[\frac{1}{3}  \left(\frac{Z_2^2}{M_2} -\frac{Z_1^2}{M_1}\right) + \left(\frac{1}{M_2}-\frac{1}{M_1} \right)  \frac{Z_1 Z_2 }{q^3 d^3} (\sin q d - q d \cos q d )\right] \left[\frac{\overline{\delta \jv}_\perp(\omega)}{\bar{n}}-\delta \Vv _{c\perp}(\omega)\right]\\
&-   \frac{8}{ M_c \omega} \int d q   \left[ \chi^h(q,\omega)-\chi^h(q,0)\right]  
\left[\frac{1}{3} \left( \frac{M_1 Z_2^2}{M_2 } + \frac{M_2 Z_1^2}{M_1 }\right)   - 2 \frac{Z_1 Z_2 }{q^3 d^3} (\sin q d - q d \cos q d ) \right] \delta \Vv _{r\perp}(\omega).
\end{split}
\tag{\ref{vcr2perp}$'$}
\label{pperp}
\end{equation}

\section{Natural frequencies and damping}
\label{SecV}

From Eq.~\eqref{ppar}  we can determine the natural frequencies of vibrations and their attenuation in Fig.~\ref{rectparr}.
Similarly, from Eq.~\eqref{pperp} we determine the attenuation of the rotation in Fig.~\ref{rectperpr}. 
We note that, during the period of the stabilization, only small frequencies $\omega$ in the spectrum of the pulse contribute.
We, therefore, make the substitution
\begin{equation}
\left[ \chi^h(q,\omega)-\chi^h(q,0)\right] /\omega \to \left. \frac{\pa\chi^h(q,\omega) }{\pa \omega}\right|_{\omega=0},
\end{equation}
rewriting Eqs.~\eqref{ppar} and \eqref{pperp} as
\begin{equation}
\begin{split}
&-\omega  \delta \Vv_{r\|}(\omega)
=  - \frac{4\pi   \omega}{  \omega_p^2}  \left( \frac{Z_2}{M_2}-\frac{Z_1}{M_1} \right) \overline{\delta \jv}_\| (\omega)  \\
&+   8 \int  d q  \left. \frac{\pa\chi^h(q,\omega) }{\pa \omega}\right|_{\omega=0}
\left\{\frac{1}{3} \left(\frac{Z_2^2}{M_2} -\frac{Z_1^2}{M_1}\right) + \left(\frac{1}{M_2}-\frac{1}{M_1} \right)  \frac{Z_1 Z_2 }{q^3 d^3}  [2 q d \cos q d+(q^2 d^2-2) \sin q d ] \right\} \left[\frac{\overline{\delta \jv}_\|(\omega)}{\bar{n}}-\delta \Vv _{c\|}(\omega)\right]\\
&-   \frac{8}{ M_c } \int d q   \left. \frac{\pa\chi^h(q,\omega) }{\pa \omega}\right|_{\omega=0}   
\left[  \frac{1}{3}  \left(\frac{M_1 Z_2^2}{M_2 } + \frac{M_2 Z_1^2}{M_1 } \right)  - 2 \frac{Z_1 Z_2 }{q^3 d^3}    [2 q d \cos q d+(q^2 d^2-2) \sin q d ]\right] \delta \Vv _{r\|}(\omega)  \\
&+    \frac{8}{ \omega} \left(\frac{1}{M_1}+\frac{1}{M_2} \right) \int  d q  \chi^h(q,0) \sin q d \frac{Z_1 Z_2}{q d} 
 \delta \Vv_{r\|}(\omega) ,
\end{split}
\end{equation}
\begin{equation}
\begin{split}
-\omega  \delta \Vv_{r\perp}(\omega)
&=  - \frac{4\pi   \omega}{  \omega_p^2}  \left( \frac{Z_2}{M_2}-\frac{Z_1}{M_1} \right) \overline{\delta \jv}_\perp (\omega)  \\
&+   8  \int  d q   \left. \frac{\pa\chi^h(q,\omega) }{\pa \omega}\right|_{\omega=0}
\left[\frac{1}{3}  \left(\frac{Z_2^2}{M_2} -\frac{Z_1^2}{M_1}\right) + \left(\frac{1}{M_2}-\frac{1}{M_1} \right)  \frac{Z_1 Z_2 }{q^3 d^3} (\sin q d - q d \cos q d )\right] \left[\frac{\overline{\delta \jv}_\perp(\omega)}{\bar{n}}-\delta \Vv _{c\perp}(\omega)\right]\\
&-   \frac{8}{ M_c } \int d q   \left. \frac{\pa\chi^h(q,\omega) }{\pa \omega}\right|_{\omega=0}
\left[\frac{1}{3} \left( \frac{M_1 Z_2^2}{M_2 } + \frac{M_2 Z_1^2}{M_1 }\right)   - 2 \frac{Z_1 Z_2 }{q^3 d^3} (\sin q d - q d \cos q d ) \right] \delta \Vv _{r\perp}(\omega).
\end{split}
\end{equation}
Performing the inverse Fourier transform, we have
\begin{equation}
\begin{split}
&-  \frac{ d^2 \delta \Rv_{r\|}(t)}{d t^2}
=   \frac{4\pi}{  \omega_p^2}  \left( \frac{Z_2}{M_2}-\frac{Z_1}{M_1} \right) \frac{ d\overline{\delta \jv}_\| (t)}{d t}  \\
&+   8 \int  d q \, \Im \! \!\left. \frac{\pa\chi^h(q,\omega) }{\pa \omega}\right|_{\omega=0}
\left\{\frac{1}{3} \left(\frac{Z_2^2}{M_2} -\frac{Z_1^2}{M_1}\right) + \left(\frac{1}{M_2}-\frac{1}{M_1} \right)  \frac{Z_1 Z_2 }{q^3 d^3}  [2 q d \cos q d+(q^2 d^2-2) \sin q d ] \right\} \left[\frac{\overline{\delta \jv}_\|(t)}{\bar{n}}-\delta \Vv _{c\|}(t)\right]\\
&-   \frac{8}{ M_c } \int d q \, \Im \! \!  \left. \frac{\pa\chi^h(q,\omega) }{\pa \omega}\right|_{\omega=0}   
\left[  \frac{1}{3}  \left(\frac{M_1 Z_2^2}{M_2 } + \frac{M_2 Z_1^2}{M_1 } \right)  - 2 \frac{Z_1 Z_2 }{q^3 d^3}    [2 q d \cos q d+(q^2 d^2-2) \sin q d ]\right] \delta \Vv _{r\|}(t)  \\
&-   8 \left(\frac{1}{M_1}+\frac{1}{M_2} \right) \int  d q  \chi^h(q,0) \sin q d \frac{Z_1 Z_2}{q d} 
 \delta \Rv_{r\|}(t) ,
\end{split}
\end{equation}
\begin{equation}
\begin{split}
- \frac{d^2 \delta \Rv_{r\perp}(t)}{d t^2}
&=  -\frac{4\pi }{  \omega_p^2}  \left( \frac{Z_2}{M_2}-\frac{Z_1}{M_1} \right) \frac{d \overline{\delta \jv}_\perp (t)}{d t}  \\
&+   8  \int  d q  \, \Im \! \! \left. \frac{\pa\chi^h(q,\omega) }{\pa \omega}\right|_{\omega=0}
\left[\frac{1}{3}  \left(\frac{Z_2^2}{M_2} -\frac{Z_1^2}{M_1}\right) + \left(\frac{1}{M_2}-\frac{1}{M_1} \right)  \frac{Z_1 Z_2 }{q^3 d^3} (\sin q d - q d \cos q d )\right] \left[\frac{\overline{\delta \jv}_\perp(t)}{\bar{n}}-\delta \Vv _{c\perp}(t)\right]\\
&-   \frac{8}{ M_c } \int d q \, \Im \! \!  \left. \frac{\pa\chi^h(q,\omega) }{\pa \omega}\right|_{\omega=0}
\left[\frac{1}{3} \left( \frac{M_1 Z_2^2}{M_2 } + \frac{M_2 Z_1^2}{M_1 }\right)   - 2 \frac{Z_1 Z_2 }{q^3 d^3} (\sin q d - q d \cos q d ) \right] \delta \Vv _{r\perp}(t).
\end{split}
\end{equation}
In the region of stabilization
\begin{equation}
\overline{\delta \jv}_{\|,\perp}(t)=\bar{n} \delta \Vv_{c \|,\perp}(t)=const.
\end{equation}
Therefore,
\begin{equation}
\begin{split}
&  \frac{ d^2 \delta \Rv_{r\|}(t)}{d t^2}
=  
   \frac{8}{ M_c } \int d q \, \Im \! \!  \left. \frac{\pa\chi^h(q,\omega) }{\pa \omega}\right|_{\omega=0}   
\left[  \frac{1}{3}  \left(\frac{M_1 Z_2^2}{M_2 } + \frac{M_2 Z_1^2}{M_1 } \right)  - 2 \frac{Z_1 Z_2 }{q^3 d^3}    [2 q d \cos q d+(q^2 d^2-2) \sin q d ]\right] \frac{d \delta \Rv _{r\|}(t)}{d t}  \\
& +  8 \left(\frac{1}{M_1}+\frac{1}{M_2} \right) \int  d q  \chi^h(q,0) \sin q d \frac{Z_1 Z_2}{q d} 
 \delta \Rv_{r\|}(t) ,
\end{split}
\end{equation}
\begin{equation}
\begin{split}
 \frac{d^2 \delta \Rv_{r\perp}(t)}{d t^2}
&=  
   \frac{8}{ M_c } \int d q \, \Im \! \!  \left. \frac{\pa\chi^h(q,\omega) }{\pa \omega}\right|_{\omega=0}
\left[\frac{1}{3} \left( \frac{M_1 Z_2^2}{M_2 } + \frac{M_2 Z_1^2}{M_1 }\right)   - 2 \frac{Z_1 Z_2 }{q^3 d^3} (\sin q d - q d \cos q d ) \right]  \frac{d \delta \Rv _{r\perp}(t)}{d t}.
\end{split}
\end{equation}
This can be written in the standard form of differential equations of self oscillations with damping
\begin{equation}
\frac{ d^2 \delta \Rv_{r\|,\perp}(t)}{d t^2}+2 \kappa_{\|,\perp} \frac{ d \delta \Rv_{r\|,\perp}(t)}{d t}+\omega_{0_\|,\perp}^2 \delta \Rv_{r\|,\perp}(t)=\0v,
\label{EQo}
\end{equation}
with
\begin{align}
&\omega_{0 \|}^2= -8 \left(\frac{1}{M_1}+\frac{1}{M_2} \right) \int  d q \, \chi^h(q,0) \sin q d \frac{Z_1 Z_2}{q d}, \label{w0par}\\
&\kappa_\|=  -\frac{4}{ M_c } \int d q \, \Im \! \!  \left. \frac{\pa\chi^h(q,\omega) }{\pa \omega}\right|_{\omega=0}   
\left[  \frac{1}{3}  \left(\frac{M_1 Z_2^2}{M_2 } + \frac{M_2 Z_1^2}{M_1 } \right)  - 2 \frac{Z_1 Z_2 }{q^3 d^3}    [2 q d \cos q d+(q^2 d^2-2) \sin q d ]\right], \label{kpar}\\
&\omega_{0 \perp}^2=0, \\
&\kappa_\perp=-\frac{4}{ M_c } \int d q \, \Im \! \!  \left. \frac{\pa\chi^h(q,\omega) }{\pa \omega}\right|_{\omega=0}
\left[\frac{1}{3} \left( \frac{M_1 Z_2^2}{M_2 } + \frac{M_2 Z_1^2}{M_1 }\right)   - 2 \frac{Z_1 Z_2 }{q^3 d^3} (\sin q d - q d \cos q d ) \right] 
\label{kperp} .
\end{align}

We note that $\kappa_\|$ and $\kappa_\perp$ are necessary positive, since
\begin{align}
&\Im \! \!  \left. \frac{\pa\chi^h(q,\omega) }{\pa \omega}\right|_{\omega=0}<0, \forall q,
\end{align}
and  expressions in square brackets in Eqs.~\eqref{kpar} and \eqref{kperp} can be shown to be positive at arbitrary values
of $M_1$, $M_2$, $Z_1$, $Z_2$, $d$, and $q$. Therefore, if $\omega_0^2\ge\kappa^2$, we have a damped oscillatory motion, which was the case in Fig.~\ref{rectparr}. Otherwise, oscillations are suppressed.

\end{document}